\documentclass{aastex62}
\usepackage{amsmath}
\usepackage{bm}
\usepackage{amsthm}
\usepackage[outdir=./]{epstopdf}
\usepackage{url}

\received{XXX}
\revised{XXX}
\accepted{XXX}
\submitjournal{ApJS (planned)}

\shorttitle{Reconstructing the Orphan Stream Progenitor}
\shortauthors{Shelton et al.}

\begin{document}

\title{An Algorithm for Reconstructing the Orphan Stream Progenitor with MilkyWay@home Volunteer Computing}

    \author{Siddhartha Shelton}
	\affiliation{Department of Physics, Applied Physics and Astronomy, Rensselaer Polytechnic Institute, 110 8$^{\rm th}$ St., Troy, NY 12180, USA}
   
   \author{Heidi Jo Newberg}
	\affiliation{Department of Physics, Applied Physics and Astronomy, Rensselaer Polytechnic Institute, 110 8$^{\rm th}$ St., Troy, NY 12180, USA}

    \author{Jake Weiss}
	\affiliation{Department of Physics, Applied Physics and Astronomy, Rensselaer Polytechnic Institute, 110 8$^{\rm th}$ St., Troy, NY 12180, USA}
    
    \author{Jacob S. Bauer}
	\affiliation{Department of Physics, Applied Physics and Astronomy, Rensselaer Polytechnic Institute, 110 8$^{\rm th}$ St., Troy, NY 12180, USA}
	\affiliation{Department of Physics, Engineering Physics and Astronomy, Queen's University, Canada}

    \author{Matthew Arsenault}
	\affiliation{Department of Physics, Applied Physics and Astronomy, Rensselaer Polytechnic Institute, 110 8$^{\rm th}$ St., Troy, NY 12180, USA}

    \author{Larry Widrow}
	\affiliation{Department of Physics, Engineering Physics and Astronomy, Queen's University, Canada}

    \author{Clayton Rayment}
	\affiliation{Department of Physics, Applied Physics and Astronomy, Rensselaer Polytechnic Institute, 110 8$^{\rm th}$ St., Troy, NY 12180, USA}

    \author{Travis Desell}
	\affiliation{Department of Software Engineering, Rochester Institute of Technology, 134 Lomb Memorial Drive, Rochester, NY 14623, USA}

    \author{Roland Judd}
	\affiliation{Department of Physics, Applied Physics and Astronomy, Rensselaer Polytechnic Institute, 110 8$^{\rm th}$ St., Troy, NY 12180, USA}
	
    \author{Malik Magdon-Ismail}
	\affiliation{Department of Computer Science, Rensselaer Polytechnic Institute, 110 8$^{\rm th}$ St., Troy, NY 12180, USA}

    \author{Eric Mendelsohn}
	\affiliation{Department of Physics, Applied Physics and Astronomy, Rensselaer Polytechnic Institute, 110 8$^{\rm th}$ St., Troy, NY 12180, USA}

    \author{Matthew Newby}
	\affiliation{Department of Physics, Temple University, 1925 N. 12$^{\rm th}$ St, Philadelphia, PA 19122, USA}
	
    \author{Colin Rice}
	\affiliation{Department of Physics, Applied Physics and Astronomy, Rensselaer Polytechnic Institute, 110 8$^{\rm th}$ St., Troy, NY 12180, USA}

    \author{Boleslaw K. Szymanski}
	\affiliation{Department of Computer Science, Rensselaer Polytechnic Institute, 110 8$^{\rm th}$ St., Troy, NY 12180, USA}
    \affiliation{Department of Physics, Applied Physics and Astronomy, Rensselaer Polytechnic Institute, 110 8$^{\rm th}$ St., Troy, NY 12180, USA}
    
    \author{Jeffery M. Thompson}
	\affiliation{Department of Physics, Applied Physics and Astronomy, Rensselaer Polytechnic Institute, 110 8$^{\rm th}$ St., Troy, NY 12180, USA}

    \author{Carlos Varela}
	\affiliation{Department of Computer Science, Rensselaer Polytechnic Institute, 110 8$^{\rm th}$ St., Troy, NY 12180, USA}

    \author{Benjamin Willett}
	\affiliation{Department of Physics, Applied Physics and Astronomy, Rensselaer Polytechnic Institute, 110 8$^{\rm th}$ St., Troy, NY 12180, USA}

    \author{Steve Ulin}
	\affiliation{Department of Physics, Applied Physics and Astronomy, Rensselaer Polytechnic Institute, 110 8$^{\rm th}$ St., Troy, NY 12180, USA}
	
    \author{Lee Newberg}

\begin{abstract}

We have developed a method for estimating the properties of the progenitor dwarf galaxy from the tidal stream of stars that were ripped from it as it fell into the Milky Way. In particular, we show that the mass and radial profile of a progenitor dwarf galaxy evolved along the orbit of the Orphan Stream, including the stellar and dark matter components, can be reconstructed from the distribution of stars in the tidal stream it produced. We use MilkyWay@home, a PetaFLOPS-scale distributed supercomputer, to optimize our dwarf galaxy parameters until we arrive at best-fit parameters.  The algorithm fits the dark matter mass, dark matter radius, stellar mass, radial profile of stars, and orbital time. The parameters are recovered even though the dark matter component extends well past the half light radius of the dwarf galaxy progenitor, proving that we are able to extract information about the dark matter halos of dwarf galaxies from the tidal debris. Our simulations assumed that the Milky Way potential, dwarf galaxy orbit, and the form of the density model for the dwarf galaxy were known exactly; more work is required to evaluate the sources of systematic error in fitting real data. This method can be used to estimate the dark matter content in dwarf galaxies without the assumption of virial equilibrium that is required to estimate the mass using line-of-sight velocities.  This demonstration is a first step towards building an infrastructure that will fit the Milky Way potential using multiple tidal streams.
\end{abstract}

\keywords{Galaxy: structure -- Galaxy: kinematics and dynamics --1
Galaxy: stellar content}


\section{Introduction} \label{sec:intro}


The stellar halo of the Milky Way galaxy is known to contain many streams of stars that have been tidally stripped from dwarf galaxies and globular clusters \citep{tidals, martin2018, mateu2018}.  The largest and best studied of these is the Sagittarius (Sgr) dwarf tidal stream (henceforth called the Sgr stream), which has been traced all the way around the Milky Way \citep{Majewski2003}.  
In addition to allowing us to glimpse recent mergers of small satellites with the Milky Way, these tidal streams can also be used to trace the density distribution of dark matter \citep{binney, Eyre&binneya, Eyre&binneyb,2018ApJ...867..101B}. Tidal stream stars are potentially very powerful for measuring the (dark matter dominated) Galactic potential because they provide information on the positions of these stars in the past;  we know that in the past the stars originated in the same Milky Way satellite.  The Milky Way's gravity is responsible for both ripping the stars from the satellite and creating the potential through which the star orbits, once it has been gravitationally stripped.

The use of tidal streams to constrain dark matter is a relatively young field, and has not yet made definitive measurements of the spatial density of dark matter, the lumpiness of the dark matter distribution, or the amount of dark matter in the progenitor dwarf galaxies.  In part, this is because we are still learning all of the factors that significantly affect the evolution of tidal streams in the Milky Way, including the bar and large satellites such as the Large Magellanic Cloud (LMC).  In addition, the fitting of streams has been plagued by large computation, which is why there has been so much effort in finding ways to fit streams without resorting to N-body simulations. 

In this paper, we present for the first time an attempt to use distributed computing provided by MilkyWay@home to solve this problem.  MilkyWay@home employs volunteered computer processors from around the world to find the the model parameters that best fit a particular dataset, given an algorithm to compute the goodness-of-fit.  In this case we optimize the mass and shape of the dwarf galaxy that disrupted to produce the observed tidal stream.  In the longer term we plan to extend the fitting procedure to multiple tidal streams simultaneously, which would make it possible to fit the shape of the Milky Way potential.  But before we describe this computationally intensive approach, we will review previous attempts to measure the shape of the Milky Way using tidal streams.

Many authors have attempted to constrain the shape of the dark matter potential using the most prominent halo stream, the Sgr stream (e.g., \citealt{Ibata2001,Majewski2003,newberg2007}).  Various studies have argued for a spherical \citep{Ibata2001,Majewski2003, Fellhauer2006}, prolate \citep{Helmi2004,Vivas2005}, or oblate \citep{Johnston2005, law2005} dark matter potential. The problem in fitting the Sgr stream stems from the fact that the trailing tidal tail favors oblate models while the leading tail favors prolate models \citep{Prior2009}. An additional complication is that both the leading and trailing tidal tails appear to be bifurcated \citep{Fellhauer2006,Koposov2012}.  

Currently, the only static model that fits all of the Sgr stream data is a triaxial halo model \citep{lawjohnston2010,Law2010}. However, this model is not favored because the Milky Way disk stars would be orbiting around the intermediate axis, and the minor axis is aligned with the direction from the Sun to the Galactic center; the stability of the disk is not guaranteed for this case \citep{Debattista2013}.  Furthermore, the axial ratios for this triaxial halo at the distances probed by the Sgr stream are considerably smaller than expected in the cold dark matter paradigm \citep{Hayashi2007,DegWidrow2013}.  

On the other hand, \cite{Willett2010} showed that if one simultaneously fit three tidal streams (Sgr, GD-1, and the Orphan Stream), with the restriction that one axis is perpendicular to the Galactic plane, the best fit triaxial halo parameters of \cite{Law2010} are recovered.  In addition, \cite{DegWidrow2013} tried to solve this conflict with $\Lambda$CDM by using a more rigorous maximum likelihood algorithm to fit 28 halo parameters to the Sgr stream using an orbit-fitting technique, and, contrary to their expectations, found a nearly identical triaxial halo model.  

\cite{Vera2013} pointed out that interaction with the Large Magellanic Cloud (LMC) would affect the orbit of the Sgr dwarf galaxy, and that this would in turn affect the tidal stream.  They were able to fit a class of models with an oblate shape in the inner galaxy ($r<10$ kpc, thus ensuring stable disk orbits), and a triaxial shape (like that of \cite{Law2010}) at large radii. The models can be made consistent with $\Lambda$CDM simulations if the potential of the LMC is included. This research opens up the possibility of a radial dependence to the average halo shape, and a time-varying component due to halo substructures (of which the largest is the LMC).  This could introduce a large number of additional parameters to the model of the Milky Way potential.  

Recently, the effect of the LMC on tidal streams has become more apparent.  \citet{koposov2019} found a misalignment of the stream direction with the streaming velocity that could not be reproduced in any static gravitational potential.  \citet{erkal2018arxiv} estimated that an LMC mass of about $1.4 \times10^{11}\,\rm{M_\odot}$ was required to produce this misalignment.  \citet{erkal2018} predicted that Tuscana III tidal stream would be perturbed by the LMC, since it passed within 15 kpc of the LMC approximately 75 Myr ago, and it is expected that many halo streams, particularly those in the south, will be affected by a very heavy LMC.

In addition to the Sgr stream, other tidal streams (Pal 5, GD-1, and the Orphan Stream) have been used in attempts to model the dark matter distribution in the Milky Way \citep{Odenkirchen2009, Willett2009, Koposov2010,newberg2010}. None of these papers reached definite conclusions, but it is clear from the attempts that different streams are sensitive to different aspects of the halo parameters. For example, streams with large eccentricity are sensitive to the mass, and streams on orbits that pass near the Galactic poles are more sensitive to the flattening. One of the reasons these papers did not reach consistent and definitive conclusions is that they did not include enough complexity.  For example, \citet{newberg2010} used the Orphan Stream orbit to find an extremely low best-fit mass of the Milky Way of $2.6 \times 10^{11} M_\odot$, integrated out to 60 kpc from the Galactic center.  We now believe this low mass was found because the Galactic model did not include the effects of the LMC, which is now thought to be much heavier than previously imagined, and has a significant affect on the stream orbit.

Many measurements have used the technique of fitting orbits to the positions and line-of-sight velocities of stars in tidal tails, under the assumption that the tidal streams followed the orbit \citep{Ibata2001,Helmi2004,Willett2010,Koposov2010,DegWidrow2013,Vera2013}. In some cases, authors disregarded the distance to the tidal stream, as the streams deviate more from the orbit in distance than in line-of-sight velocity. In most cases, N-body simulations, in which a dwarf galaxy composed of N bodies was evolved through a fixed potential, were run with the final fit parameters to demonstrate that it was a reasonable fit. Some fits varied the N-body simulations themselves in an attempt to find the best solution (e.g.- \citealt{Johnston2005,law2005,Law2010}).

New techniques for determining the Galactic potential from tidal streams are under development.  \cite{binney} showed that given the position (l, b) and line-of-sight velocity at some position along a stream, and a fixed Galactic potential, energy conservation fixes the values of the stream distance and proper motions at that position. Therefore, knowledge of the proper motions and/or distances can be used to constrain the Galactic potential. \cite{Penarrubia2012} introduced the Minimum Entropy Method (MEM), which may be used to constrain the Milky Way potential, or to test alternative theories of gravity. Using MEM, it is not necessary to evolve through a potential, or measure detailed properties of individual streams (allowing the use of very low surface brightness tidal debris). Since the energy will be distributed tightly around that of the progenitor object, and therefore has a low entropy, a technique that varies the potential (or gravity) until the entropy of the system is minimized will recover the underlying potential (or gravitational laws). \cite{Sanderson2013} propose a technique for using multiple `shells,' which form when a dwarf galaxy on a radial orbit is completely disrupted as it passes through the Galactic center, to constrain the Milky Way's gravitational potential. The shell forms at the outer turn-around radius of the debris. \cite{Price2013} propose an algorithm called ``Rewinder,'' which takes stars in a stellar stream back in time through a fixed potential, and varies the potential parameters until the stars end up in the same place, in the progenitor dwarf galaxy. This algorithm requires only a small number of stream stars to be successful, but, like many of these other algorithms, requires well-determined 6D phase space information.  Recently, \citet{bonaca2018} showed that simultaneous fitting of multiple cold streams could constrain parameters describing the Milky Way potential (which can be used to determine the spatial density of dark matter) at approximately the percent level.

The effects of halo substructures on tidal streams have also been studied by many authors \citep{Siegal2008, Quinn2008, Yoon2011}, who showed that a substructure would deflect stars so as to widen the tidal stream, or kick stars out of a tidal stream, leaving a gap.  Based on counting the gaps in the available data for four streams (three in the Milky Way, and one in M31), \cite{Carlberg2012} showed that the derived number of sub-halos is roughly consistent with $\Lambda$CDM predictions. However, the uncertainties are rather large, owing to the lack of streams with well-measured density profiles with which to carry out such a study.  Improved analysis of the stream from the globular cluster Pal 5 found \citep{CarlbergGrillmair2012} that the number of gaps in the Pal 5 stream within 23 degrees of the cluster predict a total of $\sim$1000 sub-halos more massive than $~10^6M_\odot$ within a radius of 30 kpc in the Milky Way.  The number of observed dwarf galaxies in the volume is much smaller, supporting the idea that the majority of the substructures in the Galactic halo do not have stars associated with them.  However, \citet{banik2019} showed that the observed structure of Pal 5 could be explained entirely by the effects of the Galactic bar, spiral structure, giant molecular clouds (GMCs), and globular clusters, thus alleviating the need for dark subhalos.

Most measurements of the dark matter mass of dwarf galaxies come from the velocity dispersion of the stars in that galaxy.  If the dwarf galaxy is in equilibrium then the total mass can be derived from the square of the velocity dispersion.  If only the line-of-sight dispersion is known, then the mass estimates can be inaccurate \citep{Gilmore2007, Strigari2007}, though \cite{Wolf2010} derive a method for determining accurate masses using only these line-of-sight velocities.  This works well if the mass follows the light (for example in a globular cluster system which contains no dark matter).  If the dark matter has a larger spatial extent than the stars, then the velocity dispersion of the stars measures only the dark matter inside the radius where the stars are measured, and in addition the inferred mass depends on the (unknown) radial profile of the dark matter. The exchange of energy between dark matter and baryons could change the central dark matter profile \citep{Sellwood2003, Gnedin2004, Tonini2006}.  If the centers of dwarf galaxies are baryon-dominated, then the derived dark matter content can be incorrect due to this energy exchange. Binary stars can artificially inflate the measured velocity dispersion of dwarf galaxies \citep{McConnachie2010}. 

We present an algorithm that can measure the dark matter content of dwarf galaxies without the need to assume equilibrium. The algorithm uses N-body simulations to model current stellar density distributions of a tidal stream, recovering parameters of the progenitor that inform us of the matter composition and radius distribution of both baryonic and dark matter. This method requires only (l,b) data of tracer stars from the tidal stream. As we will show, we are  able to recover parameters used to create a simulated tidal stream that is designed to be similar to the Orphan Stream. 

In this paper, we present the preliminary results of constraining the dark matter content of dwarf galaxies using N-body simulations. We will first describe our algorithm and the underlying theoretical models implemented. In discussing our algorithm, we will begin with a brief description of MilkyWay@home, the powerhouse behind our optimizations. Then we will give an overview of our simulations, describing the general setting: our representation of the Milky Way galaxy, the choice of orbit on which our dwarf galaxy is placed. We follow this with a description of the dwarf galaxy model. Next, we describe how we implemented this theoretical framework in generating a simulated dwarf galaxy. We conclude this section with a description of other various parameters needed to make the simulations work. We then move onto the comparison algorithm, describing each of the three part metric used for comparison: how they work and their role in the over all comparison likelihood. 
Finally, we conclude with our results derived from optimization with MilkyWay@home.

\section{Overview}

We develop an algorithm to probe the mass and radial extent of the dark matter in dwarf galaxies that are currently in the process of merging with the Milky Way, and that have produced the observed tidal streams of stars that we observe in the Milky Way halo. Although we plan to relax these limitations in the future, we currently assume fixed Milky Way potential, and a dwarf galaxy orbit that was fit to the position and line-of-sight velocities along the tidal stream. The dwarf galaxy is generated with two separate Plummer sphere density profiles for the stars and the dark matter. This dwarf galaxy is then evolved through the Milky Way in an N-body simulation to produce a tidal stream.

This simulated tidal stream is compared to data for actual tidal streams of stars in the Milky Way halo, and the dwarf galaxy parameters are varied until the simulated and observed streams match. We optimize the orbital time and four dwarf galaxy parameters dwarf galaxy that represent the mass and radius of each of the two Plummer sphere components (stars and dark matter). This simulated dwarf galaxy is comprised of particles that interact with each other and with an external potential.  These particles are evolved through the fixed Milky Way potential for an orbital time.  At the end of the simulation we calculate quantities such as the density of stars along the stream to determine how well the simulated parameters recreate the observed data.  We are careful to only compare observable quantities; in particular since dark matter cannot be directly observed, its properties are not used in the comparison. However, we show that the `unseen' dark matter component influences the evolution of the visible matter and the formation of the tidal debris stream.

We have developed a sophisticated comparison metric which measures how good a match the simulated histogram is to the data.  This metric compares three things: the shape of a histogram of the density of stars along the stream, the total mass of stars in the portion of the stream that is being compared, and a measure of the width of the stream as a function of angle along the stream.  The width of the stream can be measured either by the velocity dispersion of stars with spectra, or by the angular width of the stream as measured from photometry.

We created simulated data for a tidal stream by starting with a simulated dwarf galaxy and then evolving it for a specific time on a particular orbit in a fixed Milky Way potential.  We will then show that we are able to recover the input dwarf galaxy parameters and simulation time using only information from the stellar component of the tidal stream.  In this demonstration, we know the exact (constant) Milky Way potential, orbit of the progenitor, and parameterized model for the dwarf galaxy progenitor, so this is an idealized case.  Future work will explore the effects of imperfect knowledge of these things, and the systemmatic effects of all of the choices we have made in designing the simulations.  In the long term, we hope to fit all of the unknowns using data from multiple tidal streams.

\section{MilkyWay@home}

Our method of probing the dark matter content of dwarf galaxies requires significant computational resources. This is because evaluation of each set of parameters in the optimization requires an N-body simulation, which can be computationally expensive. MilkyWay@home is an approximately 800 TeraFLOPS volunteer supercomputing platform with over 14,000 active volunteers at any given time. It is specially designed to optimize model parameters, given a dataset with which to compare the model, and a function that measures the goodness of fit \citep{desellThesis}. It uses the Berkeley Open Infrastructure for Network Computing (BOINC, \citealt{boinc}), which manages the sending of `work units' to volunteered computers in every country around the world. These work units run when the volunteered computers are not being used for other tasks. When a task is finished, a number that measures how well these parameters produced the above data is sent back to our local server \citep{desellThesis, weissThesis}. Each work unit includes a set of parameters for our model.

This platform was originally designed to use statistical photometric parallax to fit a density model to turnoff stars in the Milky Way's stellar halo \citep{cole2008, newby2013, newberg2013, weiss2018a}. For this original application we use a maximum likelihood algorithm to define the density model that best fits the observed spatial distribution of stars. This project has become the primary application on MilkyWay@home, and simultaneously fits 20 or more parameters. In this paper we show how the MilkyWay@home volunteer computing platform can be used to determine 4 parameters that describe the distribution of stars and dark matter in a dwarf galaxy that produced a tidal stream and one additional parameter describing the time the dwarf galaxy evolved through the Milky Way potential.

Optimization methods come from the Toolkit for Asynchronous Optimization (TAO) \citep{desellThesis}, which was originally designed specifically for MilkyWay@home.  The optimization methods include particle swarm, differential evolution, and genetic algorithms that work well in our highly asynchronous, heterogeneous environment \citep{desell2008, desell2009, weissThesis}. For our work we have used only differential evolution.   
\section{N-Body Simulations}\label{nbody sims} 
As stated above, we place a dwarf galaxy composed of N bodies in orbit around an analytic representation of the Milky Way gravitational potential. The dwarf galaxy model requires several input parameters to determine the mass and radial distribution of the baryonic (stellar) and dark matter components: the total mass and a scale radius of each component. For the baryonic component, we specify the mass ($M_B$) and scale length ($R_{B}$). We also specify two ratio parameters used to determine the mass ($M_D$) and scale length ($R_{D}$) of the dark matter component. We use ratio values in order to explore a larger range of mass and scale length values with a narrower search range. The radius ratio ($\xi_R$) and mass ratio ($\xi_M$) values are related to the dark matter parameters by:

	\begin{align}
	    R_D &= \frac{R_{B}}{\xi_R} ( 1 - \xi_R)\\
	    M_D &= \frac{M_{B}}{\xi_M} ( 1 - \xi_M),
	\end{align}
	\hspace{2em}
	\mbox{or,}
	\hspace{2em}
	\begin{align}
	    \xi_R &= \frac{R_{B}}{R_B + R_D}\\
	    \xi_M &= \frac{M_{B}}{M_B + M_D}.
	\end{align} \label{eq:ratio_conversions}
  
In order to represent the Milky Way Galaxy, we use a theoretical model comprised of three components: the bulge, disk, and halo. The model is static; it uses analytical equations and is not comprised of bodies.  Each component requires several input parameters that determine its mass and shape. Furthermore, the orbit on which the dwarf galaxy is placed is determined by a set of position and velocity coordinates that, along with the Milky Way potential, fully characterize the orbit. 
In the future we plan to simultaneously constrain the dwarf galaxy, the Milky Way potential model, and the orbital parameters. This will require us to adapt the algorithm to fit the Milky Way and orbit parameters, and will require adjusting the likelihood function to include additional data constraints such as line of sight  velocity, stream position, and multiple tidal streams. 
The Milky Way model parameters were chosen to be the same as those used to fit the orbit for the Orphan Stream in \cite{newberg2010}. We also adopt the corresponding best fit orbital parameters given by Model 5 of Table 3 in \cite{newberg2010}. Therefore, the only parameters being constrained are those of the dwarf galaxy and the time of evolution along the orbit. 

The gravitational interactions between the bodies and with the external potential are calculated using a Barnes-Hut tree algorithm. This algorithm performs the gravitational calculations much faster than the classical Newtonian direct summation method. The computation time for the tree algorithm scales as O($N$log$N$) versus the O($N^2$) for the direct summation method \citep{TreeCode}. The tree code is also implemented as a multithreaded CPU application. 

In this section we describe the Milky Way model, the choice of orbit, the dwarf galaxy model and its implementation, and the N-body simulation itself. The N-body simulation description includes the integrator, the units being used, and the determination of the timestep and softening parameter. 

\subsection{Galactic Potential}\label{sec:galactic_pots}
The Milky Way potential components consist of a Miyamoto-Nagai disk, a Hernquist bulge, and a logarithmic halo. The bulge is represented using a spherical model:

    \begin{equation}
	\Phi_{bulge} (r) = -\frac{GM_{bulge}}{r + d} ,
    \end{equation}
where d = 0.7 kpc is a scale length, $M_{bulge}$ = 9.9 $\times$ $10^{10}$ $M_{\odot}$\citep{law2005}, and $r$ is the distance from the center. To represent the disk, a Miyamoto-Nagai disk is used:

    \begin{equation}
	\Phi_{disk} (r) = -\frac{GM_{disk}}{\sqrt{ R^2 + (b + \sqrt{z^2 +c^2})^2}},
    \end{equation}
where b = 6.5 kpc and c = 0.26 kpc are scale parameters,  R is the cylindrical radius ($R=x^2 + y ^2)$ from the center, and $M_{disk}$ = 3.4 $\times$ $10^{10}$ $M_{\odot}$ \citep{law2005}. In order to represent the halo, we use a logarithmic potential:
    \begin{equation}
	\Phi_{halo} (r) =  v_0^2 \ \mbox{ln}\left(1 + \frac{r^2}{a^2}\right),
    \end{equation}
where $v_0$ = 73 kms$^{-1}$ \citep{newberg2010} is a scale velocity, and a = 12 kpc is a scale parameter \citep{law2005}, and $r$ is, again, the distance from the center. The scale velocity, as with many other halos with a similar parameter, is determined from the rotation speed of stars at large radii with respect to the Galactic center \citep{binney}. This parameter is set so that the sum of the three components of this Milky Way Galactic potential yields those rotation speeds. 
In the N-body simulations, the accelerations due to the above equations are used (see Appendix \ref{appendix: mw accels}), rather than the potentials. It was this combination of models and parameters that yielded the best fit orbit to the Orphan stream as described in \cite{newberg2010}. Since all of the parameters are fixed, the Milky Way galaxy model requires no adjustable parameters. 

\subsection{Choice of Orbit}\label{sec:orbit}
In general an orbit is defined by a position in space (3 parameters), a velocity within that space (3 more parameters), and a potential through which the particle will travel. Since the particle can be at any place along the one-dimensional orbit, the orbit can be defined by 5 parameters. Rather than start the dwarf galaxy at all possible positions, velocities, and times in the past, we would  like to choose only those positions and orbital times that return the debris to the location we currently see it, on the orbit it is currently traveling. To achieve this, at the beginning of the simulation we perform a `reverse' orbit calculation: a single body is placed at the current debris coordinates, but with the velocities negated. The orbit is then integrated for a variable evolution time (this is one of the five adjustable parameters that are fit) and the final positions and velocities recorded. 

The dwarf galaxy is then placed at this position on the backwards orbit. The velocities are again negated, and the dwarf galaxy then evolves along the forward orbital path. The orbit is defined by the coordinates $(l,b,R)=(218^\circ, 53.5^\circ, 28.6  \mbox{ kpc})$ with velocity  $(v_x, v_y, v_z)=(-156, 79, 107)$ kms$^{-1}$ \citep{newberg2010}, thus, the orbit and the approximate ending position of tidal debris is fixed. As we will see, the integration time in the forward direction will need to be adjusted slightly compared to the reverse orbit in order to put the debris in exactly the observed location.

\subsection{Dwarf Progenitor Model}\label{sec:dwarf_model}

We choose a dwarf galaxy model that can represent the full spectrum of possible mass-to-light ratios, i.e., from mass follows light as in globular clusters to heavily dark matter dominated as expected in ultrafaint galaxies. The model must be able to represent both situations by adjustment of the model parameters. This allows us the freedom to fit the data without previous knowledge of how much dark matter should be present in the progenitor.

For our model, we adopt a double Plummer potential profile \citep{hut03} - one representing dark matter and one representing baryonic matter. The Plummer model is not the most realistic profile for modeling dwarf galaxies; it was chosen for its simplicity and ease of implementation. As we will describe further later, the probability distribution function for the Plummer model is known analytically. Furthermore, the model lacks a singularity at $r=0$ and the mass enclosed quickly falls off for larger radii. In future work, we plan to implement a wider range of dwarf galaxy models. 

To derive this model, we use the single Plummer density distribution and potential \citep{plummer1911} as a starting point:

    \begin{align}\label{single plummer}
	\rho(r) = \frac{3M}{4 \pi a^3} \left( 1 + \frac{r^2}{a^2}\right)^{-5/2},\\
	\Phi(r)=- \frac{GM}{ a}\left ( 1 + \frac{r^2}{a^2} \right )^{-\frac{1}{2}},
    \end{align}
where $a$ is a scale length and $M$ is the total mass, and $r$ is some radius. This model is robust enough to be extended to a two component model by simply adding another term to the potential:

    \begin{equation}\label{double plummer potential}
	\Phi(r)=- \frac{GM_d}{ a_d}\left ( 1 + \frac{r^2}{a_d^2} \right )^{-\frac{1}{2}} - \frac{GM_b}{  a_b} \left ( 1 + \frac{r^2}{a_b^2}\right )^{-\frac{1}{2}},
    \end{equation}
where the subscripts represent the dark matter ($d$) and baryonic matter ($b$) components, $a_d$ and $a_b$ are the dark matter and baryonic matter scale lengths, respectively, and $M_d$ and $M_b$ are the total masses of the dark matter and baryonic matter components, respectively. As a result of the linearity of Equation \ref{double plummer potential}, the spatial density distribution follows from Poisson's Equation for gravity, $\nabla^2 \Psi = - 4 \pi G\rho$:

    \begin{equation}\label{double plummer density}
	\rho(r)= \frac{3M_d}{4 \pi a_d^3} \left ( 1 + \frac{r^2}{a_d^2} \right )^{-\frac{5}{2}} + \frac{3M_b}{4 \pi a_b^3} \left ( 1 + \frac{r^2}{a_b^2} \right )^{-\frac{5}{2}}.
    \end{equation}•

Monte Carlo rejection sampling \citep{Neumann51} is used to assign positions in phase space to bodies. The positions are assigned by sampling the two components separately using their single Plummer density distribution. This ensures a proper spatial distribution for each component; the two components are essentially stacked together with a common center of mass. This is allowed because the spatial distribution is a linear combination of the two components.

The stability of the simulated dwarf galaxy relies heavily on the velocity distribution. We require stable orbits for the bodies around the combined center of mass when in empty space, i.e. no external potential. Note, however, that, unlike the spatial distribution, the velocity distribution cannot be calculated by generating the velocity distribution for each component separately and then superimposing them. This is because the velocity distribution for a Plummer model is derived from a probability distribution \citep{binney, Aarseth74} which is a nonlinear function of the $\rho(r)$ and $\Phi(r)$ of the entire system. 

Assigning velocities requires the use of a probability distribution that uses the combined potential well. A probability distribution function, when integrated in phase space, produces the probability an object will be found in that area of phase space  \citep{binney}. The spherically symmetric probability distribution function we use is (as taken from \citealt{binney}):

    \begin{equation}\label{distfunc} 
	f(\varepsilon)= \frac{1}{\sqrt{8} \pi^2} \int_0^\varepsilon \frac{1}{\sqrt{\varepsilon- \Psi}} \frac{d^2 \rho}{d \Psi^2} d \Psi ,
    \end{equation}
where $\Psi(r)= - \Phi(r)$, the combined potential of our model, $\rho(r)$ is the combined density distribution of our model and the energy, $\varepsilon$, is defined as $\varepsilon= -\frac{1}{2} m v^2  - \Phi(r)$. For a single component Plummer model, the integral in Equation \ref{distfunc} has an analytic solution. However, for a two component model this integral equation must be solved numerically. This is done by expanding the derivative in the integrand in terms of $r$. The first derivative is expanded as:

    \begin{equation}\label{derivative expansion}
	\frac{d\rho}{d \Psi}= \frac{d \rho}{dr} \frac{dr}{d \Psi} = \frac{d \rho}{dr} \left ( \frac{d \Psi}{d r} \right )^{-1},
    \end{equation}
and the second derivative is expanded as:

    \begin{equation}\label{eq:second_derivative_expansion}
	\frac{d^2\rho}{d \Psi^2}= \left [ \frac{d^2\rho}{d r^2}  \left ( \frac{d \Psi}{d r} \right )^{-1} -  \left ( \frac{d \Psi}{d r} \right )^{-2} \frac{d^2\Psi}{d\\ r^2} \frac{d \rho}{dr} \right ] \left ( \frac{d \Psi}{d r} \right )^{-1}.
    \end{equation}•

The limits of integration are converted to radial limits, $r'_{lower}$ and $r'_{upper}$, by solving for the radius, $r'$, at which the potential is equal to the energy: $\Psi(r') = \varepsilon'$, where $\epsilon'$ is either of the current limits of integration, $\varepsilon'_{lower} = 0$ and $\varepsilon'_{upper} = \varepsilon$. At the lower limit, $\Psi(r'_{lower}) = 0$ when $r'_{lower} = \infty$. However, the integrand has a well defined peak and quickly decreases to zero, allowing for an easy approximation for this limit. At the upper limit, $\Psi(r'_{upper}) = \varepsilon$, and $r'_{upper}$ is found numerically by finding the root of $\Psi(r'_{upper}) - \varepsilon = 0$. 
\vspace{.2cm}
\subsubsection{Dwarf Realization}\label{sampling}
As previously mentioned, the positions and velocities are assigned using a rejection sampling technique \citep{Neumann51}. This can be done in one dimension for both, as will be described here.
As a result of the rejection sampling technique used, a large number of random numbers are needed when creating the dwarf galaxy. These numbers are determined using a random number generator (RNG) that requires a seed, which can be set manually. This seed value is fixed for all simulations in a given optimization, which in general requires thousands of work units. This is necessary because of the distributed nature of MilkyWay@home. We require all simulations with identical parameters to create identical dwarf galaxies and thus produce identical results, which can be validated by comparing the results from different volunteered computers. 

The density distribution is spherically symmetric and a function of radius only. Cartesian positions can be determined by sampling in radius, assigning each a unit vector (requiring two randomly generated angles), and then taking the components. We choose a set of radii by rejection sampling \citep{Neumann51} under the curve

    \begin{equation}\label{sampling mass enclosed}
	\frac{dM(r)}{dr} = 4\pi r^2 \rho(r) .
    \end{equation}
This is done by accepting a radius, $r$, if the condition $ {r^2 \rho}/{(r^2\rho)_{max}} \: > \: u$, where $u$ is a random value between [0,1], is met. 
The value $(r^2\rho)_{max}$ can be found analytically for each component. We choose to make the dwarf galaxy isotropic in space. Therefore, the particle position vector, $\bm{x}$, is determined by multiplying the radius by a random unit vector, $\bm{\hat{u}}$, $\bm{x} = |r| \bm{\hat{u}}$. The unit vector is chosen by first selecting a vector, $\bm{u}= (u_1, u_2, u_3)$, created from three random numbers between [-1,1], such that $u_1^2 + u_2^2 + u_3^2 <1$. Then, $\bm{\hat{u}} = {\bm{u}}/{\sqrt{u_1^2 + u_2^2 + u_3^2}}$, is a unit vector that is randomly sampled over the surface of a unit sphere.

As shown previously, the distribution function can be reduced to a function of radius, $r$, and velocity, $v$, by expansion of the derivatives and through the calculation of energy. Since the spatial location has already been determined, we need only generate the magnitude of the particle velocity and three random numbers specifying the direction of the velocity unit vector. We choose the magnitude of the velocity using the spherically symmetric distribution \citep{binney}, 

    \begin{equation}\label{sampling velocity enclosed}
	\frac{dN(v)}{dv} = 4\pi v^2 f(v) ,
    \end{equation}
and the same rejection sampling technique as before. In this case, $(v^2f(v))_{max}$ for the given radius is found numerically. The velocity vector is then determined in the same way as the position, $\bm{v} = |v|\bm{\hat{u}}$, but using a different randomly oriented unit vector. 

The number of baryonic and dark matter bodies is fixed for all sets of simulation parameters: half of the bodies are assigned to baryons and the other half to dark matter. The mass of each component is divided evenly between their respective bodies. Currently, 20,000 bodies are used in the simulation, with 10,000 bodies assigned to each component. Assigning bodies in this way allows for fixing the number of bodies to a computationally reasonable value while still maintaining enough bodies in each component to properly represent the potential well of that component and of the entire model.

We tested our dwarf galaxy creation algorithm by monitoring the stability of the simulated dwarf galaxy in empty space by examining its radial profile, velocity, and virial ratio. By stability, we mean that it retains, without significant deviation, the initial spatial and velocity distributions, during an N-body simulation. The tests were performed for an evolution time of 4 gigayears, over which it was found that the dwarf galaxy remained stable and in virial equilibrium.

\subsection{Running the Simulation}\label{algorithm realization}
\subsubsection{N-body Integrator}\label{integrator}
The equations of motion for each body must be solved in order to determine their final positions in phase space. The only force present in the simulation is gravity. The force calculations between the bodies use a Barnes-Hut tree algorithm as described in \cite{TreeCode} and detailed in \cite{TreeCode2}. The accelerations between a body and the Milky Way potential use analytic equations which are described in Appendix \ref{appendix: mw accels}. A Velocity Verlet algorithm \citep{verlet} is used to integrate the bodies in the simulation. At the start of the simulation, the force on a body is determined and used to find the current acceleration, $a_t$. The position at the next timestep, $x_{t+1}$, is found by using the current acceleration and velocity, $v_t$:

	\begin{align}
	    x_{t+1, i} &= x_{t, i} + v_{t, i} + a_{t, i} \left( \frac{\tau^2}{2} \right ),\\
	\end{align} \label{verlet_x}
where $\tau$ is the length of the timestep, to be described later. Using this new position, the acceleration at the next timestep, $a_{t+1}$, is determined. The velocity at the next timestep is then found using a combination of the previous acceleration and the new one:

	\begin{align}
	    v_{t+1, i} &= v_{t, i} + \left ( a_{t, i} + a_{t+1, i} \right ) \left( \frac{\tau}{2} \right ).\\
	\end{align} \label{verlet_v}
Together, equations \ref{verlet_x} and \ref{verlet_v} represent the Velocity Verlet method in its original form.
\subsubsection{Simulation Units}\label{sim units}
The interaction between the bodies involves the gravitational constant, which is extremely small. To avoid floating point error, we use kiloparsecs (kpc) for distances, gigayears (Gyr) for time, and kiloparsecs per gigayear (kpc/Gyr) for velocity. For mass, we use simulation (structural) units which are derived from units where we set the gravitational constant equal to 1. Each mass unit, then, has `units' of:

    \begin{equation}
	[m] = \frac{[\mbox{kpc}^3]}{[\mbox{Gyr}^2]} = \mbox{222288.47} M_\odot.
    \end{equation}

\subsubsection{Timestep}\label{timestep}
The forces a body experiences during a simulation can be large depending on its location in the overall potential well. The timestep length must be small enough to resolve the motion of the body under the influences of such forces, but must be large enough to not introduce errors due to numerical precision, and for the simulation to be computationally feasible; the smaller the timestep length the more calculations that are needed.  In general, the timestep, $tau$, must be a small fraction of the freefall time:
    \begin{equation}\label{eq:1 comp timestep}
	\tau = \frac{1}{\gamma} \sqrt{ \frac{1}{\rho}} = \frac{1}{\gamma} \sqrt{ \frac{4\pi}{3} \frac{a^3}{M}},
    \end{equation}
where $a$ is the scale radius, $M$ the total mass in simulation units, and $\gamma$ is a tunable parameter. This rule is adapted for the two component model by calculating two parameters, $\tau_1$ and $\tau_2$:

    \begin{equation}
	\tau_1 = \frac{1}{\gamma} \sqrt{ \frac{4\pi}{3} \frac{a_{b}^3}{(M_b + M_{encl,d}) }},
    \end{equation}
and
    \begin{equation}
	\tau_2 = \frac{1}{\gamma} \sqrt{ \frac{4\pi}{3} \frac{ a_{d}^3}{(M_d + M_{encl,b}) }},
    \end{equation}
where $M_d$, $M_b$, $a_d$, and $a_b$ are defined in section \ref{sec:dwarf_model}, $M_{encl,d}$ is the dark matter mass enclosed within the baryonic scale radius, and $M_{encl,b}$ is the baryonic mass enclosed within the dark matter scale radius. The smaller of the two parameters is used as the timestep length. For the tuning parameter, we use $\gamma$ = 1000, which was found to give good simulation precision without making the computation unreasonably long. With the units for mass as described in section \ref{sim units}, the timestep has units of Gyr.

\subsubsection{Softening Length}\label{softening parameter}
The bodies in the simulation each have the gravitational potential well of a point particle, lending to strong interactions when bodies get close to each other. These strong interactions would occur with much lower frequency in a real dwarf galaxy, which has many more, much smaller particles. In addition, the strong interaction might not be resolved by the timestep in the simulation. To reduce the effects of strong interactions, the force is `softened' by removing the singularity at $r=0$ of Newton's law of gravity, thus truncating the bodies' individual potential wells. This is done by adding a `softening parameter' to the distance between bodies, turning each body into a hard sphere. The force between two bodies is then: 
    \begin{equation}
	  F_{ij} = \frac{GM_iM_j}{ r_{ij}^2 + \epsilon^2}.
    \end{equation}
For a single component model, $\epsilon$ is determined by the following value:

    \begin{equation}
	\epsilon^2 = \frac{a^2}{\beta N},
    \end{equation}
where $\beta$ is a tunable parameter (we use $\beta=100$) used to reduce the softening length, N is the number of bodies in the simulation, and $a$ is the scale length. For a two component model we have two scale lengths, so we use a `center of mass' scale length determined by:

    \begin{equation}
	a = \frac{ M_b a_b + M_d a_d}{M_b + M_d},
    \end{equation}
where $M_d$, $M_b$, $a_d$, and $a_b$ are defined in section \ref{sec:dwarf_model}.


\section{Metric for Comparing Simulation Results to Data}
We require a method to compare the simulated tidal stream distribution to actual tidal debris data drawn from astronomical surveys. Since the dark matter in tidal streams cannot be observed, we compare the observed stars with only the stellar component of the simulated tidal stream. The {\it likelihood} that these two distributions are the same guides our search for dwarf galaxy parameters; each set of initial parameters produces a different density distribution along the stream and thus a different likelihood. The similarity of the actual and simulated tidal streams is of primary concern. Our comparison method is described in this section. 

The metric used for comparison is comprised of three components, each of which accounts for different physical aspects of the stream. The first component focuses strictly on the density of stars as a function of angle along the stream, as depicted in a histogram. The second focuses on the stellar mass that each data set represents. The final component looks at a histogram of the width of the stream as a function of angle along the stream.

\subsection{Creating Histograms of Density and Stream Width Along the Stream}
As we have said previously, the final state of the simulation is represented in a histogram. The first step in creating a histogram is to remove the dark matter bodies from the distribution. Next, the coordinates of each remaining body is transformed to solar-centered angular coordinates, with the equator of the system aligned along the tidal debris stream under investigation. We use ($\Lambda$, $\beta$) coordinates, and transformation rules, as defined for the Orphan Stream in \cite{newberg2010}, where $\Lambda$ is parallel to the stream and $\beta$ is perpendicular to the stream. A region of the sky is selected in which the bodies will be analyzed. This region can be narrow or encompass the entire sky. Normally, the region is kept narrow in $\beta$ to avoid capturing multiple wraps of the stream, if there are any, and to avoid particles that may no longer be associated with the stream. The region in $\Lambda$ is kept wide for simulated data, encompassing almost the entire sky. Depending on the availability of stellar data, the range is narrowed to avoid including bins without any observational information. 

In this region, data from both the observations and the simulation are binned in $\Lambda$, to represent the density of stars as a function of angular position along the stream. The total number of bodies that fall within the entire region, $N_T$, is also noted. The count and error in each bin is given by $N_i \pm \sqrt{N}$. After the binning is finished, the count in each bin is normalized so that the sum of all the histogram bins is one: 

    \begin{equation}
	\frac{N_i}{N_T} \pm \left\{
		\begin{array}{ll}
		     \frac{\sqrt{N_i}}{N_T} \indent \mbox{if } N_i \neq 0 \\
		     \frac{1}{N_T}          \hspace{.5em} \indent \mbox{if } N_i = 0
		\end{array}.
	    \right.
    \end{equation}

We also characterize the width of the stream as a function of $\Lambda$. We have tried two different widths, the line of sight velocity dispersion and the dispersion in the $\beta$ coordinate. We have shown that either is successful for constraining the progenitor mass, but we currently use the $\beta$ dispersion because the observational data for this quantity is more readily available. The algorithm can handle either, along with their respective errors.

The dispersion in some quantity, $\alpha$, whether it be line of sight velocity or the $\beta$ coordinates, is given by:

    \begin{equation}\label{disp_original}
	\sigma^2 = \frac{1}{N - 1} \sum_i^N (\alpha_i - \bar{\alpha})^2,
    \end{equation}
where $\alpha_i$ is the quantity for body $i$, $\bar{\alpha}$ is the average of that quantity for the entire bin, and $N$ is the number of bodies in the bin.
This can be written as:

    \begin{equation}\label{disp_comp_simp}
	\sigma_\alpha^2 = \sum_i \frac{\alpha_i^2}{N-1} - \left (\frac{N}{N-1} \right ) \left ( \sum_i \frac{\alpha_i}{N} \right )^2.
    \end{equation}

Outliers are rejected until all bodies greater than 2.5$\sigma$ from the average quantity, $\bar{\alpha}$, for that bin are removed. The average is then recalculated with the remaining bodies. The intrinsic change to the $\alpha$ distribution involved in outlier rejection is accounted for by multiplying $\sigma_{\alpha}$ by $\sqrt{1.11}$ to correct for the reduction in dispersion caused by the removal of the tails of the Gaussian distribution. 

We perform six rounds of outlier rejection, which was found to be enough to remove the most extreme outliers; it is computationally quicker to reject outliers six times rather than testing to determine whether the algorithm has converged. After this process is complete, the error in the final dispersion value is also calculated. For the simulation, this value can be derived analytically and is given by

    \begin{equation}\label{eq:disp_error}
	\sigma \pm \delta_{\sigma} = \sigma \pm \sigma \sqrt{\frac{N + 1}{N-1}\frac{1}{N}} ,
    \end{equation}
where $N$ is the remaining number of bodies not rejected as outliers. Therefore, for each bin, there is both a calculation of the line of sight velocity dispersion and $\beta$ dispersion, both with outliers rejected, and with their respective theoretical errors.

\subsection{Stellar Density Component of the Likelihood}
We compare the two stellar density distributions using an Earth Mover Distance (EMD, \citealt{emd}) technique. This method is preferable to a bin-by-bin comparison such as the $\chi^2$ method because it avoids situations where small differences between the histograms result in huge differences in the goodness of fit. For example consider two streams, one evolved for some time, $T$, and another evolved for slightly longer, $T+\Delta t$, where $\Delta t$ is just large enough for the progenitor core (in this case much more densely populated than the surrounding stream) to be counted in a different bin. Although these two streams would be overall very similar, performing a bin by bin comparison would yield a poor likelihood value because there is no benefit to the highly populated bins being close to each other.  Although the peak of the likelihood is in the correct place using a bin-by-bin algorithm, the likelihood surface would be difficult to navigate because parameters that are closer to the correct parameters will not necessarily produce likelihoods that are close to the best likelihood.

An Earth Mover Distance (EMD) calculation resolves this issue by providing a method that compares the overall shape of the two histograms instead of a bin-by-bin comparison. The method essentially asks how much deformation has to be applied to one histogram to arrive at the other.  The name is derived from the example of moving dirt around to make one pile of dirt look like the other  In order to avoid the need to create or destroy ``dirt" in order to make one pile of dirt look like the other, we normalize the two histograms before calculating the EMD. 

The EMD is calculated from a transportation problem which determines the minimum amount of work needed to perform this deformation. If $\rho_{sim}= \left\lbrace  (\Lambda_1, N_1), ... , (\Lambda_m, N_m) \right\rbrace$ are the bin centers and counts in one histogram, and $\rho_{data}= \left\lbrace  (\Lambda_1, N_1), ... , (\Lambda_n, N_n) \right\rbrace$ are the bin centers and counts in the other histogram, the `work' to perform the deformation is given by:
    \begin{equation}
	 W(\rho_{sim}, \rho_{data}) = \sum_{i=1}^m \sum_{j=1}^n d_{ij}f_{ij} ,
    \end{equation}
where $f_{ij}$ is a flow matrix and $d_{ij}$ is a matrix of the ground distances between bins \citep{emd}.  In this case, the ground distance is the angle between the $i^{\rm th}$ bin in the simulated histogram and the $j^{\rm th}$ bin in the data histogram.  The flow matrix element, $f_{ij}$, is the number of (normalized) bodies in the $j^{\rm th}$ bin of the data histogram that have to be moved to the $i^{\rm th}$ bin of the simulated histogram to perform the deformation.  The flow matrix is subject to the constraints that: (1) you must move a positive number of bodies out of each bin in the data histogram, (2) you cannot move more bodies out of a data bin that there are in the bin to begin with, and (3) you must reproduce the number of bodies present in each bin of the simulated histogram when you are finished.  
    
The EMD algorithm finds the flow matrix that minimizes the work, $ W(\rho_{sim}, \rho_{data})$. The EMD value is defined in terms of the optimal flow matrix and the ground distance matrix as follows:
    \begin{equation}
	EMD(\rho_{sim}, \rho_{data}) = \frac{\sum_{i=1}^m \sum_{j=1}^n d_{ij}f_{ij} }{\sum_{i=1}^m \sum_{j=1}^n f_{ij}}.
    \end{equation}
In our case, since we normalize both histograms so that the sum of the bins is 1, the sum of the elements of the flow matrix (the denominator in the calculation of the EMD) is also one, and the EMD is equal to the work.  The method for finding the optimal flow matrix is explained in detail in \cite{emd}. The probability returned from the EMD calculation is scaled and returned as the stellar density portion of the likelihood value,
    \begin{equation}
	P_{shape} = 
	    \begin{cases}
		1 - \frac{EMD}{EMD_{max}} \indent &0 < EMD < EMD_{max}\\
		0  &otherwise
	    \end{cases},
    \end{equation}
where $EMD_{max}$ is a scaling factor so that for reasonable configurations of bodies, $EMD/EMD_{max}$ ranges from zero to one.

We normalized the two histograms being compared so that we did not need a ``source'' or ``sink'' of bodies. However, by doing this, information on the number of bodies, and subsequently the stellar mass represented in each histogram, was lost. To penalize the likelihood when the total mass of stars does not match the total mass of the stellar bodies, we use a mass cost function. While EMD compares the shapes of the two histograms, the mass cost compares the total stellar mass each histogram represents.

\subsection{Mass Cost Component of the Likelihood}
We introduce a penalty to the likelihood if the mass represented by the data histogram differs from the mass represented by the simulation histogram. Note that each body in the simulation represents a fixed mass of stars, $M_{sim}$, equal to the total mass of the baryonic component divided by the number of baryonic bodies. Similarly, each observed star in the data histogram represents some mass of stars, as traced by the type of star that is observed. 
The mass of each ``star'' in the data histogram is inflated by how many solar masses each tracer star represents, in solar masses. The total mass in the simulated histogram is $M_{sim}N_{sim}$, and the total stellar mass represented by the data histogram is $M_{data}N_{data}$. We then determine the number of standard deviations, $\sigma$, by which these two integrated mass estimates differ. $N_\sigma$ is found by dividing the difference between the two masses by the sum of the two errors, added in quadrature.

The errors in the masses represented by each histogram can be determined by considering how the two histograms are made. While they are both counting experiments with two outcomes (an object is either included or not), they differ in that the simulation has a finite number of objects that can be included in the histogram, while the number of stars in the stellar stream is not exactly known but is known to be large.  The sample of stars in the data histogram are tracer stars that represent a larger stellar population, just as each body in the simulation is assigned a mass that represents many solar masses. The total mass in each histogram is determined by the mass represented by each object (tracer star or body) and the total number of histogrammed objects. The error in the total object count is the primary source of error considered in this calculation, though there could be an error in our knowledge of the stellar mass represented by each tracer star. 

Because the number of bodies in the simulation is fixed, and the number in the part of the sky that is ``observed'' is selected from this maximum, the error in the counts for a simulation histogram is given by binomial statistics:

    \begin{equation}
	N_{sim} \pm \sqrt{ N_{sim}\left( \frac{N_{sim}}{N_{total}} \right ) \left ( 1 - \frac{N_{sim}}{N_{total}} \right )},
    \end{equation}
where $N_{sim}$ is the number of bodies in the simulation histogram, and $N_{total}$ is the number of bodies in the simulation (baryonic component), i.e., the limit in the number of bodies that can be in the histogram. 

The error in the counts for a histogram made from data is given by Poisson statistics, and will depend on the method for producing the data histogram as well as the background, if any. For a simple case without background, 

    \begin{equation}
	N_{data} \pm \sqrt{ N_{data}  } .
    \end{equation}

Therefore, the number of $\sigma$ difference between the masses represented by the two histograms is given by,
    \begin{equation}
	N_{\sigma} = \frac{  M_{sim}N_{sim} - M_{data}N_{data}} {\sqrt{ M_{data}^2 N_{data} + M_{sim}^2N_{sim}\left( \frac{N_{sim}}{N_{total}} \right ) \left ( 1 - \frac{N_{sim}}{N_{total}} \right )} }.
    \end{equation}

This equation is then plugged into a probability density describing whether the mass within the two histograms are different by chance, given by:

    \begin{equation}\label{eq:massprob}
	P_{mass} = e^{-\frac{N_{\sigma}^2}{2}},
    \end{equation}
The ``most similar'' that the two histograms can be is identical, with a $N_\sigma=0$. The maximum probability is one, giving it the same scale as the EMD component, by construction.

\subsection{$\beta$ Dispersion Component of the Likelihood}
In addition to using the density of stars along the stream to constrain the model parameters, we also use the stream width. 
Historically, streams have been separated by likely dwarf galaxy and likely globular cluster progenitors by the width of the tidal stream produced, suggesting this width might be a powerful discriminant of progenitor mass. 

At first, we tried using a 2D Earth Mover Distance method to compare the density distribution in the $\Lambda$ and $\beta$ directions, but this was time-consuming and did not work when the parameters caused the stream to shift in the $\beta$ direction. Shifts in the $\beta$ direction could indicate that the orbital parameters need to be adjusted, but we do not currently optimize the properties of the progenitor and the orbit of the progenitor simultaneously.

We then tried a histogram of the line-of-sight velocity dispersion, which worked well for simulated data but failed on the real data because the observational velocity errors were too large and the number of stars with spectra too small to measure the parameters of the Orphan Stream progenitor. Therefore, histograms of spatial dispersion in the $\beta$ coordinate, binned in the same $\Lambda$ bins as the number density used for the EMD, are used to compare the data with the simulations.

Similarly to the cost component, we calculate a $\sigma$ difference between the width histograms. We calculate the $N_\sigma$ between individual bins between the two histograms, using the probability density given by Equation \ref{eq:massprob}. The final probability density is a multiplicative series of each bin's probability density, 

    \begin{equation}
	P = \prod_{i} P_i = \prod_{i} e^{-\frac{N_{\sigma, i}^2}{2}} = e^{-\frac{1}{2} \left ( N_{\sigma, 1}^2 +  N_{\sigma, 2}^2 + ... + N_{\sigma, m}^2 \right )},
    \end{equation}
where the sum in the exponent is given by
    \begin{equation}
	N_{\sigma}^2 = \sum_i^{N_{bins}} \frac{ (\sigma_{\beta,data,i} - \sigma_{\beta,sim,i})^2}{(\delta_{\sigma_{\beta,data,i}}^2 + \delta_{\sigma_{\beta,sim,i}}^2)}.
    \end{equation}
The $\delta$ values are the errors in the $\sigma_\beta$ values. For the simulated histograms, we have width information for every bin. This is not the case for the stellar histogram. Therefore, only bins with width information are included in this sum. The probability density for the width is then,
    \begin{equation}
	P_{width} = e^{-\frac{N_{\sigma}^2}{2}}.
    \end{equation}
As with the stellar density and mass cost components, the $\beta$ component varies between $0$ and $1$.

\subsection{Final Likelihood}
The likelihood $\mathcal{L}$ provides a metric with which to measure the similarity of the two histograms and thus the stellar distributions they represent. Together, the three components described above allow for a comparison of the tidal debris stream represented by two sets of histograms. Each component is combined into a final probability given by their product:
    \begin{equation}
	\mathcal{L} = \prod_{i} P_i.
    \end{equation}
Or, in log space:
    \begin{equation}
	\ln \left[ \mathcal{L} \right] = \sum_{i} \ln (P_i ) = \ln  (P_{shape})+ \ln (P_{mass} ) + \ln (P_{width} ).
    \end{equation}
We note that we have normalized the individual terms in the probability density distribution so that they range from zero to one. The maximum probability for each component in log space is zero; for identical histograms, ln$(\mathcal{L})=0$. Any different normalization would produce additive constants in log space. Therefore, normalizing the component probability densities amounts to a shift in the likelihood surface and not a change in the shape. We prefer to keep the maximum likelihood at zero. Our model parameters are constrained by finding the parameter set for which the likelihood, $\mathcal{L}$, is a maximum.

\subsection{Best Likelihood Determination: Reducing Chaotic Behavior}
Our method for navigating the likelihood surface to find the best-fit parameters relies on the assumption that small changes in the initial parameters produce small changes in the resulting histograms, so that the likelihood surface is fairly smooth. Unfortunately, chaos can contradict this assumption, and in particular we found that small changes to the initial conditions, including changing the random seed, can change the final position of the dwarf galaxy along the orbit, and therefore cause an unacceptably large change in the likelihood. If the forward evolve time is held constant, the location of the peak of the distribution of stars would fluctuate back and forth along $\Lambda$ as each dwarf galaxy parameter is changed, making it difficult to find the peak likelihood.

As described in previous sections, the optimizer currently fits five parameters: four dwarf parameters and the evolution time.  One particle is moved backwards along the orbit for the evolution time, and we run the simulation in the forward direction for as long as it takes for the progenitor to get to the current position, as determined by the time to get to the maximum likelihood. At the beginning of the simulation, the number of time steps in the simulation is determined based on the timestep length and the simulation time. 
At some percentage of the total number of time steps, and for every subsequent time step, the simulation will create a histogram from the current state of the simulation during that time step, compare with the input histogram and calculate a likelihood. We save the best likelihood value from all of the steps tested, along with the associated histogram. The percentage value is an input parameter, and is currently set to 98 percent, so the program searches for the maximum likelihood for a window of forward evolution times between 98\% and 102\% of the evolution time parameter.

\section{Optimization with MilkyWay@Home}

For the N-body application, MilkyWay@home optimizes over five parameters. It uses a differential evolution algorithm specially modified to work in a distributed supercomputer \citep{desellThesis, weissThesis}. This optimizer creates and maintains a population of the best simulation likelihoods. It replaces members of the population with likelihoods that are better as they are returned from volunteer machines. The current population is used to determine what values to send out to volunteers to try. A complete description of the method and algorithm can be found in Chapters 2 and 5 of \cite{desellThesis}. Chapter 2 of \cite{desellThesis} describes the basic algorithm in its normal form and Chapter 5 describes the implementation of this algorithm in our highly asynchronous environment. 

We use 20,000 bodies in our simulations, of which 10,000 represent the stars and 10,000 represent the dark matter.  The CPU time required to run each simulation varies wildly based on the timestep, which depends strongly on the central density of the the progenitor dwarf galaxy.  The time to return a likelihood result from an individual volunteer varies from a few minutes to a few days depending on the model parameters being simulated and the volunteer machine.  We typically evaluate 50,000 likelihood calls to search the parameter space before the optimization converges. Full convergence is reached when returned values no longer differ, and the members of the population have converged to be identical.  Optimization on Milkway@home takes about two to three weeks to fully converge.  This time does not increase when several optimizations are performed simultaneously.

The differential evolution algorithm has several parameters which affect how quickly our search converges. The algorithm uses a population of parents made up of a parameter set and an associated likelihood value. For N-body, we use a population size of 50. Children are generated from two randomly selected parents. The crossover probability is set to 0.9 and the differential scaling factor is set to 0.8. For a detailed explanation of the roles of these parameters in the algorithm see Section 4 of \cite{weiss2018a}.

\section{Testing the Algorithm with Simulated Data}
To test our ability to recover parameters for our model, we created a simulated data set with known parameters, and attempted to recover the parameters used in its creation using MilkyWay@home. Table \ref{table:sim_fits} shows the parameters used to generate the simulated (``correct'') data histogram, as well as the range (``search range'') over which the optimization algorithm was allowed to search.  The search range for the mass ratio parameter ($\xi_M$) was set as wide as possible, with a mass to light ratio ranging between 1000:1 to 1:1, representing dwarf galaxies that are strongly dark matter dominated to dwarf galaxies with no dark matter at all. The baryonic mass ($M_B$) search range was also set fairly wide compared with the dwarf galaxy mass used in \cite{newberg2010} of 2.5$\times10^6$ M$_\odot$. We again use the static Milky Way potential and orbit from \cite{newberg2010}, as described in Sections \ref{sec:galactic_pots} and \ref{sec:orbit}, respectively.

The baryonic mass search range of 1 to 100 structural units corresponds to a mass of 2.2 $\times10^5$ M$_\odot$ to 2.2 $\times10^7$ M$_\odot$, respectively. The search range for the baryonic scale radius ($R_B$) was narrower, between 0.05 kpc to 0.5 kpc. The smaller range was chosen because any lower would mean a very dense object, requiring a very small timestep and thus an unfeasibly long computational time on the average client computer. The upper end of the range limits the scale radius to a relatively large Milky Way dwarf galaxy, though the LMC, SMC, Sgr, Crater II and Canis Major are all larger. The radius ratio ($\xi_R$) search range also goes from 0.1 to 0.6, corresponding to a dark matter halo that is 100 times more extensive as the baryonic component to mass follows light. The simulation time was also given a fairly wide search range corresponding to the dwarf galaxy core remaining undisrupted to it being mostly disrupted as seen in Figure \ref{fig:partial_disrupt}.
    \begin{figure*}[!ht]
	\centerline{\includegraphics[width = 10cm]{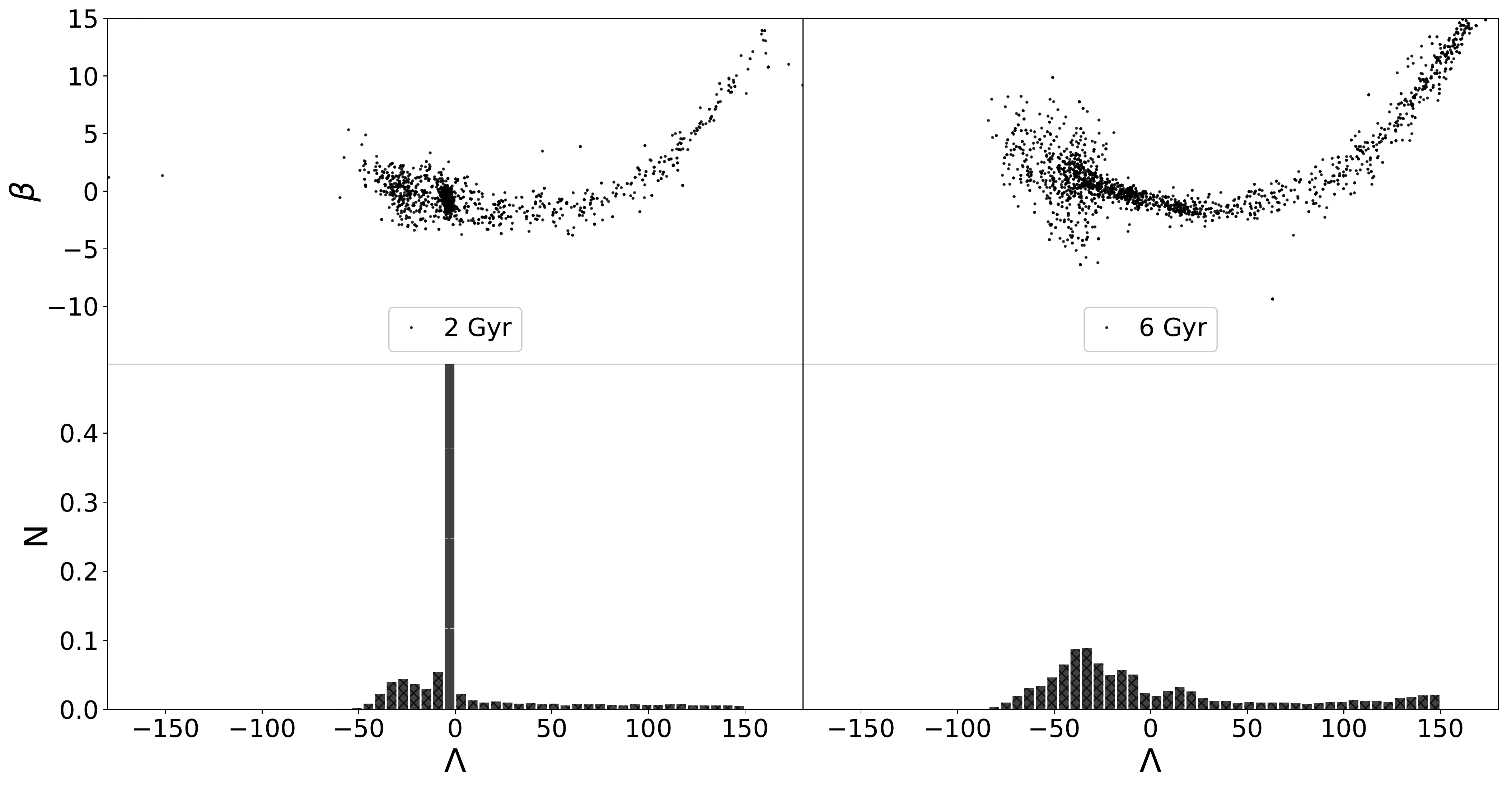}}
	\caption{The stellar portion of a dwarf galaxy with the parameters used for the simulated data set evolved for 2 Gyr (left) and 6 Gyr (right), with their respective histograms below.  The center of the stream strays from $\beta=0$ because the stream does not follow a great circle on the sky, as viewed from the position of the Sun.  A longer evolution time produces a longer tidal stream.  In 6 Gyr, the dwarf galaxy has completely dissolved.}
    \end{figure*}\label{fig:partial_disrupt}

\begin{figure*}[!ht]
	\centerline{\includegraphics[width = 10cm]{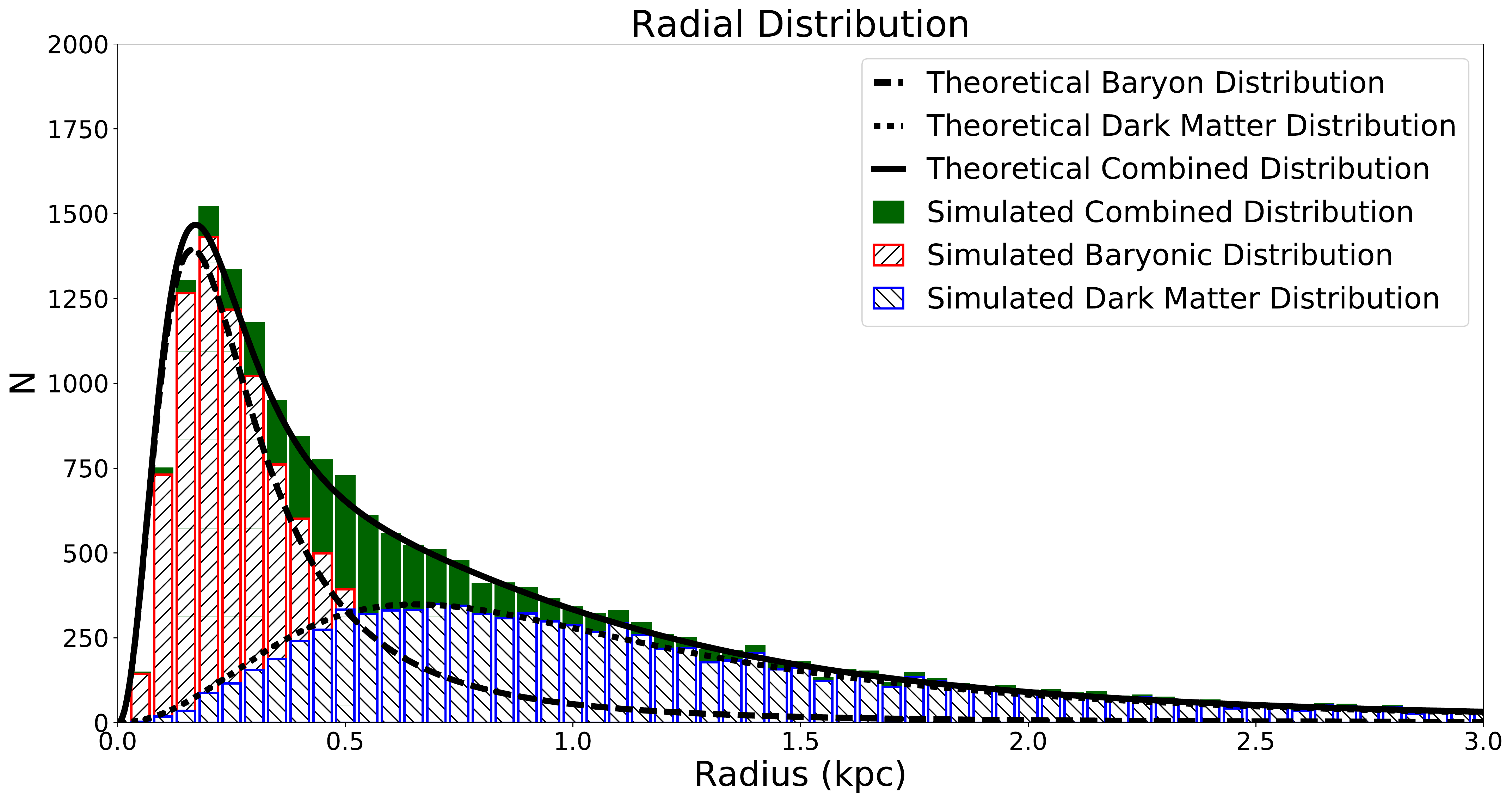}}
	\caption{The radial distribution of the simulated dwarf galaxy model used in the test optimization. The dwarf galaxy comprises of a baryonic matter core, of scale radius 0.2 kpc, with an extended dark matter component of scale radius 0.8 kpc.}
	\label{fig:mixed_radial}
    \end{figure*}

\begin{center}
	\begin{figure}[!ht]
	\centerline{\includegraphics[width = 13cm]{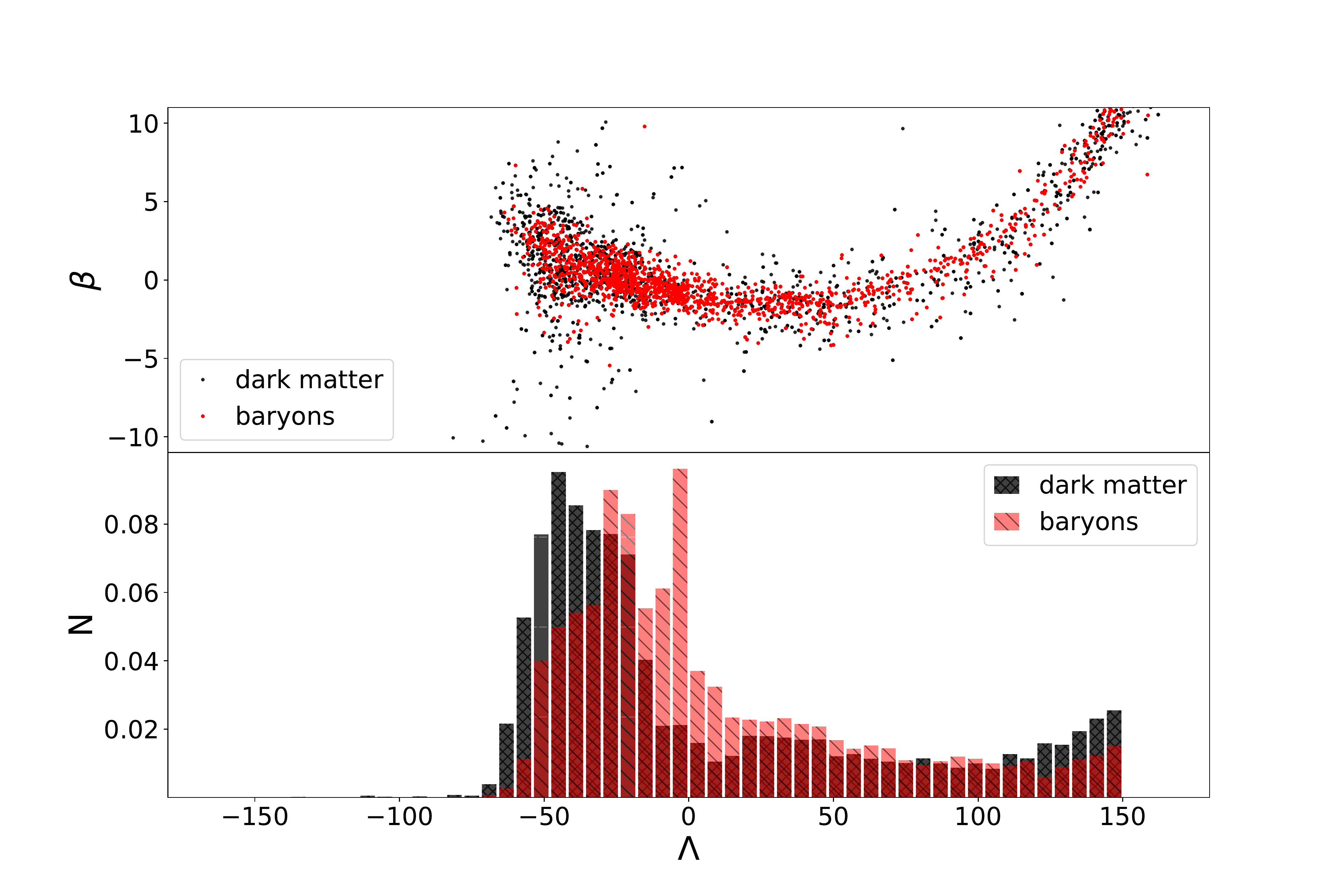}}
	\caption{The simulated data histogram. Above is a sub-sampled $\Lambda, \beta$ plot of the two components. Below are the corresponding density distribution histograms. As can be seen in either panel, the two components have similar density distributions. The main difference is the baryonic core present at $\Lambda\sim0$. The stream becomes more distant at negative $\Lambda$, causing the bodies to pile up in angular position in the sky.}
	\label{fig:simulated_data}
	\end{figure}
    \end{center}

We use a different seed in the creation of our simulated data set from that used for the optimization runs. For the baryonic component parameters, we use similar parameters as those of the single component Plummer model used in the N-body simulations in \cite{newberg2010}, which is approximately 12 simulation units, or 2.7 $\times10^6$ M$_\odot$, with a scale length of 0.2 kpc. For the dark matter component, we use a mass to light ratio of 4:1, with the dark matter radial extent also four times as large as the baryonic component. This represents a dwarf galaxy whose stellar component is embedded in an extended dark matter halo. This can be seen in Figure \ref{fig:mixed_radial}, which shows the radial distribution of both components and the entire dwarf galaxy as a whole. The mass ratio was also set to 0.2, which sets the dark matter mass to four times as massive as the baryonic component, at 48 simulation units, or $1.1 \times 10^7$ M$_\odot$. These parameters were found by fixing the baryonic component and altering the dark matter component until the final distribution, shown in Figure \ref{fig:simulated_data}, is reminiscent of the Orphan stream stellar distribution given in Figure 5 of \cite{newberg2010}. The likelihood value obtained from comparing a histogram made from these parameters with a simulation also run with these parameters but with the random number seed used in the optimization (different from the one used in the creation of the histogram) is given along with the ``correct'' parameters in Table \ref{table:sim_fits}.
A ($\Lambda, \beta$) plot of this dwarf galaxy evolved for 3.95 Gyr with histograms of both components is shown in Figure \ref{fig:simulated_data}.


     \clearpage
    \begin{center}
	\begin{table}[!ht]
	\centering
	    \begin{tabular}{|c|c|c|c|c|c|c|}
		\hline
		Parameters& Evolve Time & $R_B$ & $\xi_R$ & $M_B$ &$\xi_M$&Likelihood\\
		\hline \hline
		Correct & 3.95 	& 0.2  	 &  0.2   & 12.0   & 0.2& -30.265\\ 
		\hline
		Search Range & [2.0 - 6.0] & [0.05 - 0.5]  & [0.1 - 0.6]  & [1.0 - 100.0]  & [0.001 - 0.95]    &\\
		\hline
		Trial 1 &  4.175 $\pm$ 0.014& 0.199 $\pm$ 3.162 $\times10^{-5}$ & 0.216 $\pm$0.008 & 12.070 $\pm$ 0.096& 0.226 $\pm$ 0.012&-13.905 \\  
		\hline
		Trial 2 & 4.117 $\pm$ 0.127&  0.198 $\pm$ 3.162 $\times10^{-5}$ & 0.220 $\pm$ 0.008& 12.111 $\pm$ 0.066 & 0.239 $\pm$ 0.014& -14.163\\
		\hline
		Trial 3& 4.089 $\pm$ 0.052 & 0.196 $\pm$ 3.162$\times10^{-5}$ & 0.211 $\pm$ 0.008& 12.045 $\pm$ 0.035& 0.221 $\pm$ 0.011& -14.337\\  
		\hline
	    \end{tabular}
	\caption{Best fit values for optimizations on histograms made from a different seed as the dwarf galaxy.}
	\label{table:sim_fits}
	\end{table}
    \end{center}
    \begin{center}
	\begin{table}[!ht]
	    \centering
	    \begin{tabular}{|c|c|c|c|c|c|}
		\hline
		Parameters& Evolve Time (Gy) & $R_B$ (kpc) & $R_D$ (kpc) & $M_B$ ($M_\odot$)& $M_D$ ($M_\odot$)\\
		\hline \hline
		Correct 		& 3.95 	& 0.2  	 &  0.8  &2.667$\times10^{6}$&1.067$\times10^{7}$\\ 
		\hline
		Trial 1 &  4.175 $\pm$ 0.014& 0.199 $\pm$ 3.162 $\times10^{-5}$&  0.724 $\pm$ 0.035 & 2.683 $\times10^{6}$ $\pm$ 2.1339 $\times10^{4}$& .916 $\times10^{7}$ $\pm$ 6.652 $\times10^{5}$\\
		\hline
		Trial 2 & 4.117 $\pm$ 0.127&  0.198 $\pm$ 3.162 $\times10^{-5}$  & 0.702 $\pm$ 0.034 & 2.692$ \times10^{6}$ $\pm$  1.4671 $\times10^{4}$ & .859 $\times10^{7}$ $\pm$ 7.246 $\times10^{5}$\\
		\hline
		Trial 3 & 4.089 $\pm$ 0.052 & 0.196 $\pm$ 3.162$\times10^{-5}$& 0.732 $\pm$ 0.035 & 2.677$ \times10^{6}$ $\pm$ .7780 $\times10^{4}$ & .942 $\times10^{7}$ $\pm$ 6.315 $\times10^{5}$\\
		\hline
	    \end{tabular}
	    \caption{The fitted values as in Table \ref{table:sim_fits}, but converted to more physical quantities.}
	    \label{table:sim_fits_stellarvalues}
	\end{table}
    \end{center}

    \begin{center}
	\begin{figure}[!ht]
	\centerline{\includegraphics[width = 10cm]{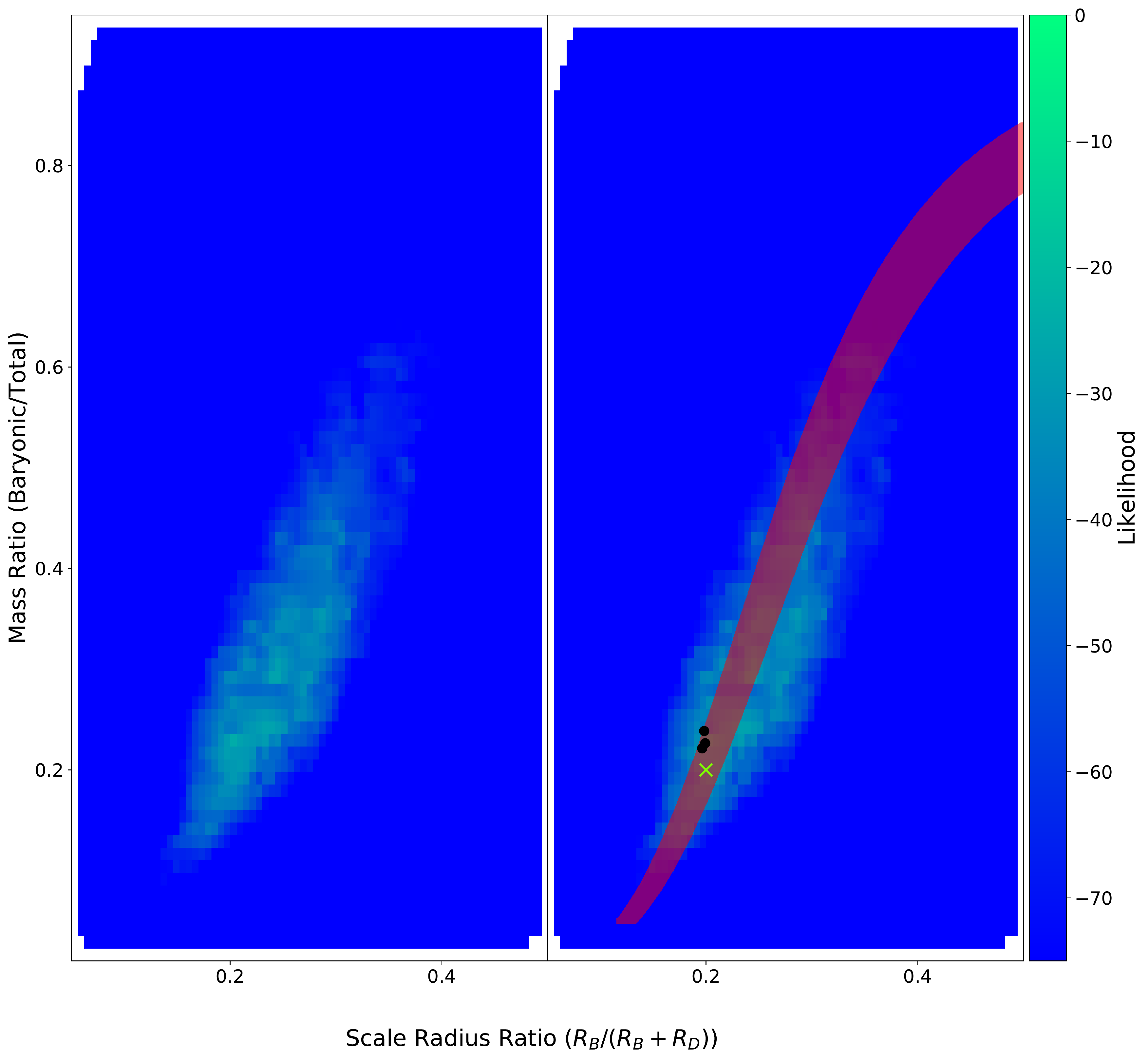}}
	\caption{Likelihood surface of the two ratio parameters, the radius ratio and mass ratio, that are used to determine the dark matter scale radius and mass. The right plot is the same as the left but also shows the fitted values as black dots, and the correct answer as a green cross. Also plotted is the region, displayed in red, in these two parameters where the dark matter mass within the half light radius is approximately the same as the simulated progenitor. As the heat map shows, there is a ridge in the surface that corresponds fairly well with the red region. However, the ridge is peaked at the dark matter mass and scale length of the simulated progenitor, indicating that information about these two quantities is present in the resulting stream.}
	\label{fig:heatmap}
	\end{figure}
    \end{center}
    
    \clearpage

We used a wide search range of $\xi_M =$ [0.001, 0.95] for the mass ratio. The dark matter mass represented by this ratio is dependent on the baryonic mass, as given by Equation \ref{eq:ratio_conversions}, which has a search range of $M_B$ = [1, 100] simulation units, or [2.2$\times$10$^5$ M$_\odot$, 2.2 $\times$10$^7$ M$_\odot$]. Looking at all possible configurations of baryon mass and dark matter mass accessible during the optimization, the mass of the dark matter can be found anywhere between 10$^4$ M$_\odot$ and 2.2 $\times$10$^{10}$ M$_\odot$. Looking at the narrowest search range possible, that which would occur if the mass of the baryons was fixed at the correct value, the accessible values of dark matter masses would still range from 1.4$\times$10$^{5}$ M$_\odot$ to  2.67$\times$10$^{9}$ M$_\odot$. This is an incredibly large range of masses. 

The dark matter scale radius was also searched over a wide range.  The radius ratio was searched in the range from $\xi_R$ = [0.1, 0.6]. The dark matter scale radius represented by this ratio is dependent on the baryonic scale radius, which has a range of $R_B$ = [0.05, 0.5] kpc. Again looking at the possible configurations of baryonic scale radius and dark matter scale radius accessible during the optimization, the scale length of the dark matter can be set anywhere between 0.033 kpc to 4.5 kpc. Looking at the narrowest range possible, if the baryonic scale radius was fixed to the correct answer, the accessible values of dark matter scale lengths would range from 0.33 kpc to 1.8 kpc.

Three identical but independent optimizations were performed for this histogram. The results of the three optimization trials are shown in Table \ref{table:sim_fits}. We have converted the parameters to more physical parameters (rather than ratios) in Table \ref{table:sim_fits_stellarvalues}. In each case we were able to recover the parameters used in the creation of the dwarf model. As the table shows, the values for the baryonic mass are all within 1\% of the correct answer. The baryonic scale radius is also recovered quite well, with all of the fitted values within 2\% of the correct answer. This is not surprising since the likelihood metric uses the baryonic component in the comparison. During the optimization baryonic components parameters are usually the first ones fit, indicating that the algorithm is very sensitive to them. The evolution time was also recovered, with the fitted values within 5\% of the correct value.  We have also recovered the dark matter parameters, with the mass within 20\% of the correct value and the radius within 10\% of the correct value. We performed a preliminary error analysis using a Hessian matrix method. These values are shown in \ref{table:sim_fits} in their original form and propagated for table \ref{table:sim_fits_stellarvalues}, using standard error propagation techniques.

The fact that we are able to recover the dark matter parameters is important because the radial extent of the dark matter component extends well beyond the half light radius (see Figure \ref{fig:mixed_radial}). In fact, with these parameters, most of the dark matter mass is located outside of the half light radius of 0.261 kpc, with 1.43 simulation mass units (3.2 $\times10^5$ M$_\odot$), of dark matter mass enclosed within. This means our algorithm was able to find the total dark matter mass within 20\%, even though less than 3\% of the dark matter mass was within the half light radius.  

Evidence that the properties of the dark matter in the progenitor can be ascertained from the distribution of stars in the stellar stream it produces is shown in Figure \ref{fig:heatmap}. This figure shows a two dimensional likelihood surface generated by varying the two ratio parameters that determine the dark matter properties, while fixing the other parameters, and recording the resulting likelihood when compared with a histogram made from the correct parameters (but different random seed). The left figure shows this surface, with a ridge clearly visible where the recorded likelihood is much better than the surrounding parameter space. On the right is the same plot but with the correct answer indicated by a green `x', and, as we will discuss further later, the fitted values from the optimization algorithm are shown as black dots. In addition, the figure on the right has an overlay that indicates the region of parameter space in which the dark matter mass within the half-light radius is approximately constant. Note that this region aligns with the ridge in the likelihood surface, but that the likelihood surface is not constant along the ridge. The likelihood is higher near the mass and radial profile for which the simulated data was created. If the algorithm was sensitive exclusively to the mass within the half light radius, the fitted values would have been randomly distributed along the shaded region. The fitted values align very closely to the correct value, indicating an implied relationship between the values that the algorithm is fitting and the mass outside of the half-light radius, or at a minimum the shape of the dark matter distribution near the center. 

\begin{figure}[!ht]
	\centering
	    \includegraphics[width = 16cm]{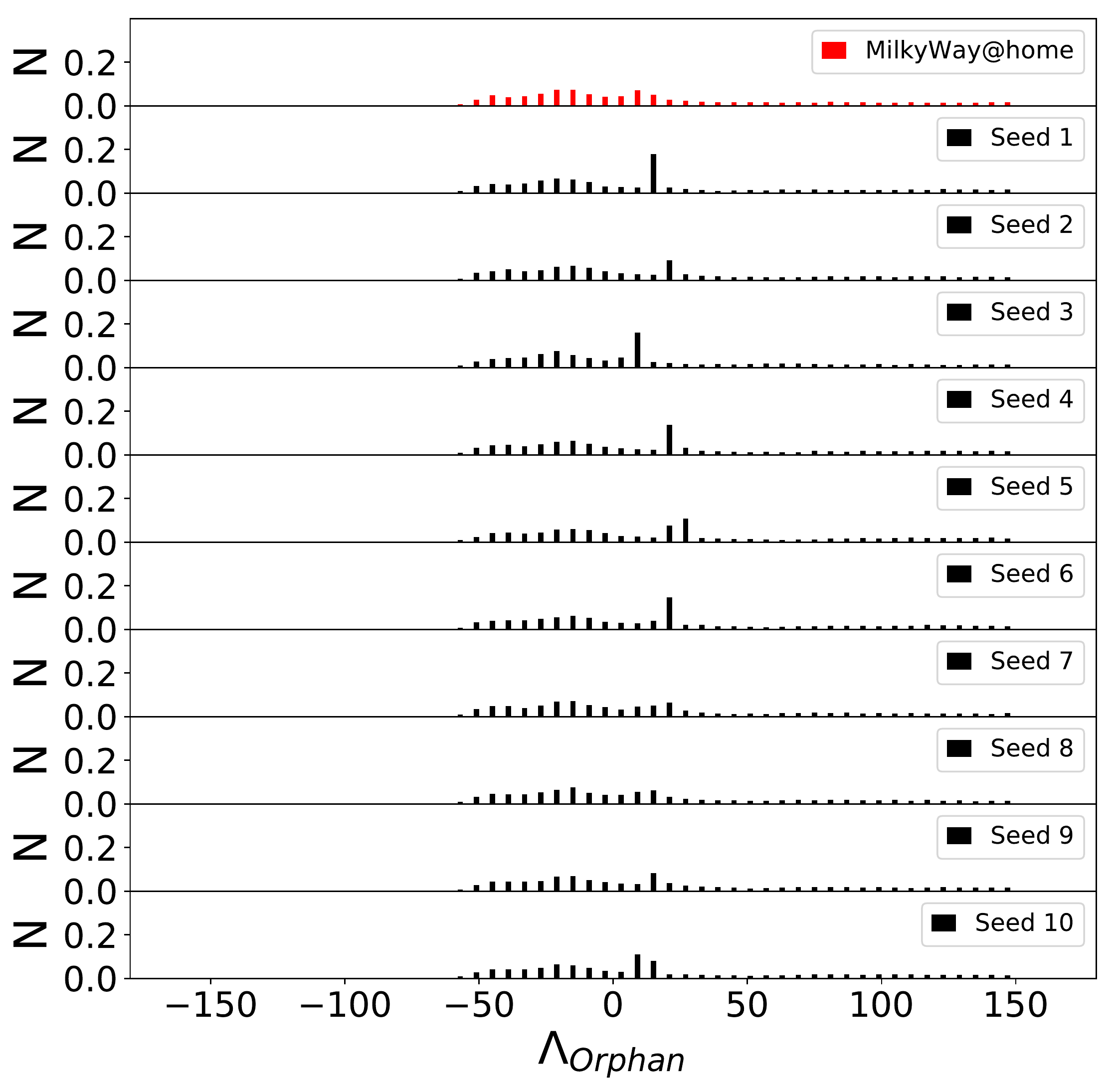}
	\caption{Plots of the simulation histogram used in the optimization, made with 10 different seeds. For some choice of seeds, the main peak at $\Lambda_{Orphan} \sim$ 25$^o$ is less pronounced. As the plot shows, there is also some fluctuations in the location and shape of the peak seen at $\Lambda_{Orphan} \sim$ 25$^o$.}
	\label{fig:diffseed_hists_lmbda}
    \end{figure}

The goodness of fit is strongly dependent on the seed used to create the simulated data and the seed used in the optimization. For some choices of random number seed, some of the distinctive characteristics of the simulated data histogram, such as peaks, can disappear. 
The simulated data histogram  (seed 5 in Figure \ref{fig:diffseed_hists_lmbda}) is one such case. Our particular choice of random seed, seed 3 in the figure, produces a tidal stream that is similar to the real data in that it has a strong peak (see Figures \ref{fig:diffseed_hists_lmbda}). 
The best fit parameters from MilkyWay@home, using a different seed, therefore may not reproduce the data. The dependence on the seed deserves a more thorough analysis in the future.

These results show a proof of concept: our distributed supercomputer, MilkyWay@home, using the described optimization algorithm, can recover the best fit parameters of a progenitor dwarf model, given sufficient photometric data for tracer stars in the Orphan Stream, and that the model for the dwarf galaxy, the Milky Way potential, and the orbital parameters are known exactly. Importantly, using our N-body method, we are able to extract information about the dark matter distribution well outside of the half light radius. This can be compared to other methods of fitting tidal stream properties, such as streakline \citep{kutter2012}, are limited to information contained within the tidal radius.
Tests will be required to determine how sensitive we are to errors in these quantities. In principle, the algorithm can be extended to fit these as well, with additional data required to constrain the additional parameters. These results give us confidence that we can fit accurate dwarf galaxy parameters to a histogram of stellar stream data: the parameters that MilkyWay@home fits are reasonable estimates of the parameters of the dwarf galaxy progenitor for the Orphan Stream.


\section{Discussion}
We have been able to recover parameters used in the creation of a simulated tidal stream. This demonstrates our algorithm's ability to traverse the complicated likelihood surface of our five parameters.  The fact that we are able to recover these parameters when we have used a different seed for the creation our simulated data set shows that we are able to extract four parameters for the dwarf galaxy progenitor from the tidal stream that is produced, including two parameters for the dark matter distribution. We must emphasize that this was done in the ideal case where the form of the dwarf galaxy, its orbit around the Milky Way, and the time-independent Milky Way potential were known exactly. 

We place the dwarf galaxy in orbit around a static representation of the Milky Way galaxy. Such a placement assumes the evolution of the Milky Way galaxy and the Universe has already taken place. In reality, dwarf galaxies form and disrupt in a universe that is constantly evolving. Of particular concern is the effect of the LMC, which has been shown to have a significant effect on the Orphan Stream, in particular \citep{erkal2018arxiv}.  At a minimum, this time-varying potential need to be included.

In the future, the algorithm will be expanded to simultaneously fit the dwarf galaxy, orbit, a more physical dwarf galaxy progenitor model, and the Milky Way parameters. We believe we can simultaneously fit the parameters in the orbit because the orbit was originally fit using the line-of-sight velocities, distances, and sky positions of stream stars, and these constraints were not used to constrain the dwarf galaxy parameters.  It is hoped that simultaneous fitting of multiple tidal streams will constrain the parameters of the Milky Way, and that the effects of large satellites can be included by making better measurements of the individual satellites in question. It is not known how many parameters in the dwarf galaxy progenitor model we will be able to fit.

We used a Hessian matrix in our determination the statistical error in the fitted dwarf galaxy parameters.  However, we have done little to evaluate the systematic error in these values.  The most obvious sources of systematic error are imperfect knowledge of the Milky Way potential, progenitor orbit, and the form of the progenitor model (for example it could be rotating and include a disk).  In determining the systematic error to our fits, we will also need to explore the effect of each parameter set throughout the algorithm. These include the best likelihood starting point, the number of iterations used in rejection sampling, the effect of the orbital parameters and the form of the Milky Way potential, the timestep calculation, and softening parameter, among others. We will need to alter each parameter to determine their effect on the likelihood values.  We plan to explore the potential sources of systematic error in future work.

On effect that we have considered is the possibility that the sudden placement of the dwarf galaxy into the Milky Way potential, wherever the backwards orbit happens to end, can cause tidal shocks in dwarf galaxy. In order to ascertain whether placing the dwarf galaxy in this way affects the orbit we allowed the dwarf galaxy to relax in place before evolving. It was found that there was no significant difference in the tidal stream produced by the dwarf galaxy when it was placed in orbit around the Milky Way with and without relaxation. We also placed the dwarf galaxy in orbit around a Milky Way potential which was ramped over time from empty space to the full potential model described in section \ref{sec:galactic_pots}. The potential was ramped at the very beginning of the simulation while the dwarf galaxy was allowed to orbit and while it was held in place. It was also found that there was no significant effect on the disruption of the dwarf galaxy or its orbital path when the ramping was performed over a few million years. From these two results it was determined that placing the dwarf galaxy around a fully formed Milky Way potential did not adversely affect the simulation; tidal shocks, if present, did not have a noticeable affect on the dwarf galaxy orbit. Neither did assuming a fully formed Milky Way galaxy. 

We plan to run this algorithm on real data for the Orphan Stream to constrain the properties of its progenitor.  This requires measurements of the density and width of the tidal stream as a function of angle along the stream, and an estimate for the stellar mass represented by each tracer star.  This might require that we include the LMC's (time-dependent) gravity in the simulation.

\section{Conclusions}
We have developed a method of fitting N-body simulations to tidal debris, that is effective in constraining the shape and mass of the dwarf galaxy progenitor from which it formed. Our dwarf galaxy model is comprised of two components that separately represent the baryonic and dark matter components.  We also estimate the time the dwarf galaxy has been in its present orbit. Our model provides a straightforward way of probing the dark matter content of these galaxies without the assumption of virial equilibrium.

We present a metric for comparing our simulated tidal stream distribution to that compiled from stellar data, and show that it is effective by using it to optimize N-body simulation parameters to match a simulated data set. The comparison metric is comprised of three components, each constraining a different aspect of the tidal stream: the shape of the stellar density along the stream, the total stellar mass in the stream, and the width of the stream as a function of angle along the stream. The best fit parameters for the N-body model are determined using a differential evolution method on MilkyWay@home.

We generated a tidal stream in a static Milky Way potential for a dwarf galaxy with a stellar mass of $2.7 \times 10^6 M_\odot$, a stellar scale radius of $0.2$ kpc, a dark matter mass of $1.067 \times 10^7 M_\odot$, and a dark matter scale radius of $0.8$ kpc.  The M/L ratio for this progenitor is $\sim 5$.
We were able to recover the stellar mass and scale length parameters for a simulated progenitor dwarf galaxy to 1\% and 2\%, respectively.  We were able to recover the dark matter mass and scale length parameters to 20\% and 10\%, respectively, even though less than 3\% of the dark matter mass was within the half light radius. 

In the future we plan to expand the algorithm to simultaneously fit the dwarf galaxy, orbit, a more physical dwarf galaxy progenitor model, and the (possibly time-dependent) Milky Way parameters.  This will require an extension of the metric for comparing our simulated tidal stream to actual tidal stream data.  In addition, we will explore the systematic errors introduced by inexact knowledge of each of the model components, as well as from the tunable parameters in our algorithm. We also must test whether dynamical friction has a significant effect on the disruption of the dwarf galaxy as it orbits the Milky Way. We are also exploring the affects of having several subhalos present around the Milky Way galaxy, and the affects of large structures such as the LMC on the determination of the progenitor dwarf galaxy parameters.  The algorithm will also be run with real data to constrain the Orphan Stream progenitor.

\acknowledgements
We would like to thank the more than 200,000 MilkyWay@home volunteers for their countless hours of donated computing time, their constant community involvement with debugging and development, and their financial support over the past decade. We would also like to thank and acknowledge contributions made by The Marvin Clan, Babette Josephs, Manit Limlamai, and the 2015 Crowd Funding Campaign to Support Milky Way Research. This publication is based on work supported by the National Science Foundation under grant No. AST 16-15688 and the NASA/NY Space Grant fellowship.

We use data from the Sloan Digital Sky Survey. Funding for the SDSS and SDSS-II has been provided by the Alfred P. Sloan Foundation, the Participating Institutions, the National Science Foundation, the U.S. Department of Energy, the National Aeronautics and Space Administration, the Japanese Monbukagakusho, the Max Planck Society, and the Higher Education Funding Council for England. The SDSS Web Site is http://www.sdss.org/.

The SDSS is managed by the Astrophysical Research Consortium for the Participating Institutions. The Participating Institutions are the American Museum of Natural History, Astrophysical Institute Potsdam, University of Basel, University of Cambridge, Case Western Reserve University, University of Chicago, Drexel University, Fermilab, the Institute for Advanced Study, the Japan Participation Group, Johns Hopkins University, the Joint Institute for Nuclear Astrophysics, the Kavli Institute for Particle Astrophysics and Cosmology, the Korean Scientist Group, the Chinese Academy of Sciences (LAMOST), Los Alamos National Laboratory, the Max-Planck-Institute for Astronomy (MPIA), the Max-PlanckInstitute for Astrophysics (MPA), New Mexico State University, Ohio State University, University of Pittsburgh, University of Portsmouth, Princeton University, the United States Naval Observatory, and the University of Washington.

\appendix

\section{Accelerations Due to Milky Way Potentials}\label{appendix: mw accels}
During the calculations of forces the potential equations are not used, but rather their derivatives: the accelerations due to those potentials. Furthermore, the accelerations are calculated for each component in Cartesian space. The parameters used below are the same as the parameters involved in the respective potentials. 

For the spherical model, the accelerations are given by:

    \begin{equation}
	\bm{a} =  -\frac{M \bm{x}}{r (r + d)^2}.
    \end{equation}

For the Miyamoto-Nagai disk the accelerations are given by:

    \begin{equation}
	\left ( 
	    \begin{array}{c} 
		a_x \\a_y\\a_z 
	    \end{array} 
	\right ) 
	= -\frac{M}{\left [ x^2 + y^2 + \left ( b + \sqrt{z^2 + c^2 } \right )^2 \right ]^{\frac{3}{2}}  } 
	\left ( 
	    \begin{array}{c} 
		x\\ y\\z \frac{ b + \sqrt{z^2 + c^2 } }{\sqrt{z^2 + c^2}} 
	    \end{array} 
	\right ).
    \end{equation}

As a result of the flattened nature of the cylindrical disk potential, the acceleration in the z direction is modified as indicated above.

For the log halo:
    \begin{equation}
	\left ( 
	    \begin{array}{c} 
		a_x \\a_y\\a_z 
	    \end{array} 
	\right ) 
	= -\frac{2v_0^2\gamma^2}{\gamma^2(x^2 + y^2 + a^2) + z^2} 
	\left ( 
	    \begin{array}{c} 
		x\\ y\\z/\gamma^2 
	    \end{array} 
	\right ),
    \end{equation}
where $\gamma$ is a flattening parameter. It is used to make the halo oblate or prolate. However, we use a spherically symmetric halo so we set $\gamma=1$.

\clearpage
\nocite{*}
\bibliographystyle{aasjournal}
\bibliography{research}

\end{document}